\documentstyle[12pt]{article}
  \advance\oddsidemargin  by -1in
  \advance\evensidemargin by -1in
\def\lsim{\mathrel{\rlap{\lower4pt\hbox{\hskip1pt$\sim$}}
    \raise1pt\hbox{$<$}}}         
\def\gsim{\mathrel{\rlap{\lower4pt\hbox{\hskip1pt$\sim$}}
    \raise1pt\hbox{$>$}}}         

\def\be{\begin{equation}}
\def\ee{\end{equation}}
\def\bq{\begin{eqnarray}}
\def\eq{\end{eqnarray}}
\def\bm{\boldmath}

\begin{document}
\pagestyle{empty}

\vspace{2.0cm}
 
\begin{center}
 
{\large \bm \bf  Gluon Spin in a Quark Model \\}

\vspace{1.0cm}

{\large V.~Barone$^{a,}$\footnote{Also at II
Facolt{\`a} di Scienze MFN, 15100 Alessandria, Italy.},
T.~Calarco$^{b}$ and A.~Drago$^{b}$ \\}

\vspace{1.0cm} 

{\it $^{a}$Dipartimento di
Fisica Teorica, Universit\`a di Torino\\
and INFN, Sezione di
Torino,    10125 Torino, Italy \medskip\\
$^{b}$Dipartimento di Fisica, Universit{\`a} di Ferrara \\
and INFN, Sezione di Ferrara, 44100 Ferrara, Italy \medskip\\ }

\vspace{1.0cm}

{\large \bf Abstract \bigskip\\ }

\end{center}
 
We compute the gluon polarization $\Delta G$ 
and the gluon total angular momentum $J_g$ 
in  the Isgur--Karl  
quark model. Taking into account self-interaction contributions,
we obtain for both quantities a positive value 
of about 
$0.24$ at the model scale $\mu_0^2 \simeq 0.25$ GeV$^2$. 
This estimate is reasonably close to the expectation from polarized deep 
inelastic scattering analyses. 

PACS: 13.88.+e, 14.70.Dj, 12.38.Bx, 12.39.Jh

\vfill
 
\pagebreak

\baselineskip 20 pt
\pagestyle{plain}

The EMC spin effect, {\it i.e.} the discovery that a surprisingly small 
fraction of the proton spin is carried by quarks, has been a major issue 
of hadronic physics for quite a long time by now \cite{mauro}. 
In the naive quark model one expects that quarks carry all 
proton's spin, that is $\Delta \Sigma \equiv 
\int dx \, (\Delta u + \Delta d + \Delta s) \simeq 1$. 
Taking into account relativistic effects this value 
is typically lowered to about $0.7-0.8$, still much larger 
than the experimental finding for the singlet axial charge
\cite{SMC}. 
This discrepancy is what has been termed ``the proton spin crisis''. 
The solution of the puzzle is known \cite{altarelli}. 
Due to the axial anomaly,  the singlet 
axial charge, which is the 
experimentally 
measured quantity, is not simply $\Delta \Sigma$ but 
the combination 
$a_0 = 
\Delta \Sigma - (N_f \, \alpha_s/2 \pi)\, \Delta G$, where
$\Delta G$ is the gluon helicity. A small value of 
$a_0$ ($\sim 0.3$) is thus compatible with a relatively large
value of $\Delta \Sigma$ if one allows for a sizeable 
$\Delta G$. 
It has been shown \cite{stefano} that
the experimental data suggest $\Delta G = 1.6 \pm 0.9$ 
at $Q^2 = 1$ GeV$^2$. 
The question arises whether a realistic
model of the proton -- or any non perturbative 
calculation -- can predict a 
gluon polarization of such an order of magnitude. 

The situation is rather controversial. 
In \cite{jaffe} Jaffe gave a negative answer 
to the question: he found $\Delta G \sim -0.4$ in the MIT 
bag model at a scale $\mu_0^2 \simeq 0.25$ GeV$^2$ and an 
even more negative value ($\sim -0.7$) in the non--relativistic 
quark model. On the contrary, a recent QCD sum rule calculation 
\cite{piller}
gives $\Delta G \simeq 2.1 \pm 1.0$ at 1 GeV$^2$. Hence two 
different non perturbative approaches seem to give contrasting results. 
Another open issue \cite{balitsky} is the relative weight 
of $\Delta G$ and $J_g$, the gluon total angular momentum, or
--~otherwise stated~-- the sign and the size of the gluon 
orbital angular momentum $L_g$. 

The purpose of this Letter is to clarify these 
problems. 
We shall show that: {\it i)} the negative value for 
$\Delta G$ obtained by Jaffe is a consequence of neglecting 
``self--angular momentum'' effects; {\it ii)} the quark 
model is not incompatible with a positive and sizeable 
gluon polarization; {\it iii)} a specific calculation leads 
to a value of $\Delta G$ which is not far from what is required 
by the polarized DIS data and from what is obtained in the 
QCD sum rule computations; {\it iv}) $J_g$ 
is close to $\Delta G$ at the model scale, 
and its value is in good agreement with the finding 
of Ref.~\cite{balitsky}. 
 
 In practical calculations we shall use 
the Isgur--Karl (IK) model \cite{IK}, which is sufficient for 
our purposes. We do not aim in fact at a precision 
calculation, but only at a qualitative study and  an 
order-of-magnitude estimate. 
Moreover, the possibility to carry out analytically the 
calculations allows for a better control of the result.
Our main conclusions would be unchanged 
should we use more sophisticated models of the nucleon. 

The angular momentum sum rule for the nucleon reads
\cite{ratcliffe,hoodbhoy}
\be
\frac 12=\frac 12{{\Delta\Sigma}}(Q^2)+{{L_q}}(Q^2)+
{{J_g}}(Q^2)
\label{1}
\ee
where $\Delta \Sigma$ is the quark spin, $L_q$ the quark orbital 
angular momentum and $J_g$ is the total angular momentum carried by the 
glue. All quantities on the r.h.s. of eq.~(\ref{1}) are 
gauge invariant and depend in general on a scale $\mu^2$ 
\cite{ratcliffe,hoodbhoy}. 
We shall be concerned with the gluonic term in eq.~(\ref{1}), 
which is given by
\be
J_g(\mu^2) = \langle p \uparrow \vert 
\int \, d^3 r \, \left[\vec r \times \left ( \vec E (\vec r) \times
\vec B(\vec r) \right )\right]^3 \vert p \uparrow \rangle\, .
\label{j1}
\ee
$J_g$ cannot be decomposed in a gauge invariant manner
into an orbital ($L_g$) and a spin ($\Delta G$)
part  written in terms of local fields. 
Nevertheless in the ${\cal A}^+=0$ 
gauge it is possible to give a local 
expression for  the gluon 
in terms of the color fields \cite{jaffe}
\be
{{\Delta G}}(\mu^2)=\langle p\uparrow|\int{d^3 r \,
2 \, {\rm Tr}\left\{\left[{{\vec{E}(\vec{r})\times\vec{A}(\vec{r})}}\right]^3+
{{\vec{A}_\perp(\vec{r})\cdot\vec{B}_\perp(\vec{r})}}\right\}}|p\uparrow
\rangle .
\label{2}
\ee

The renormalization of the operators appearing in eqs.~(\ref{j1}, \ref{2}) 
induces a scale dependence of $J_g$ and $\Delta G$. 
However, in a quark model  
the matrix elements in eqs.~(\ref{j1}, \ref{2}) do not depend on any scale. 
The model describes the nucleon at a fixed scale, which is 
 usually very low ($\lsim 0.5$ 
GeV$^2$). 

The color fields generated by the quarks are ($a=1,\ldots, 8$)
\begin{eqnarray*}
\vec E(\vec r) &=& \sum_{a=1}^8
\vec{E}^{{a}}({{\vec{r}}})  =  
-\vec{\nabla}\, \sum_{a=1}^8 \Phi^{{a}}({{\vec{r}}})\;;\\
\vec B(\vec r) &=&
\sum_{a=1}^8
\vec{B}^{{a}}({{\vec{r}}})  =   
\vec{\nabla}\times \, \sum_{a=1}^8 \vec{A}^{{a}}({{\vec{r}}})\;,
\end{eqnarray*}
where the potentials 
\begin{eqnarray*}
\Phi^{{a}}({{\vec{r}}}) & = & 
\frac{1}{4\pi}\sum_{{{i}}=1}^3\frac{\lambda^{{a}}_{{i}}}{2}
\int{d^3r^\prime
\frac{\rho_{{i}}(\vec{r}^{\,\prime})}
{|{{\vec{r}}}-\vec{r}^{\,\prime}|}}\;;\\
\vec{A}^{{a}}({{\vec{r}}}) & = & 
\frac{1}{4\pi}\sum_{{{i}}=1}^3\frac{\lambda^{{a}}_{{i}}}{2}
\int{d^3r^\prime
\frac{\vec{j}_{{i}}(\vec{r}^{\,\prime})}
{|{{\vec{r}}}-\vec{r}^{\,\prime}|}}
\end{eqnarray*}
are written in terms of the single--quark charge densities
\be
\rho_{{i}}({{\vec{r}}})=g\,
\delta^3({{\vec{r}}}-\vec{r}_{{i}})
\nonumber 
\ee
and current densities ($c=$ convective, $m=$ magnetic) 
\begin{eqnarray*}
\vec{j}_{{i}}({{\vec{r}}}) & = & 
\vec{j}_{{i}}^c({{\vec{r}}})+
\vec{j}_{{i}}^m({{\vec{r}}})\\
\vec{j}_{{i}}^c({{\vec{r}}}) & = & -\frac{ig}{2m}\left[
\vec{\nabla}_{\vec{r}_{{i}}}
\delta^3({{\vec{r}}}-\vec{r}_{{i}})+
\delta^3({{\vec{r}}}-\vec{r}_{{i}})
\vec{\nabla}_{\vec{r}_{{i}}}\right]\\
\vec{j}_{{i}}^m({{\vec{r}}}) & = & 
\frac{g}{4m}\vec{\nabla}_{{{\vec{r}}}}\times
\left[\vec{\sigma}_{{i}}
\delta^3({{\vec{r}}}-\vec{r}_{{i}})\right] .
\end{eqnarray*}
The expressions above are written 
in the non relativistic form, in view of their 
use 
in the IK model. 

We now make a gauge transformation to new potentials 
${\cal A}^{\mu}$, enforcing the condition ${\cal A}^{+} = 0$ 
\begin{eqnarray*}
{\cal A}^{0a}({{\vec{r}}}) & = & 
\Phi^{{a}}({{\vec{r}}})\;,\\
\vec{\cal A}^{{a}}({{\vec{r}}}) & = & 
\vec{A}^{{a}}({{\vec{r}}})-\vec{\nabla}
\int_0^z{d\zeta \,\Phi^{{a}}(x,y,\zeta)}
\;.
\end{eqnarray*}
Eqs.~(\ref{j1}, \ref{2}) become
\be
J_g(\mu^2) = \sum_{i,j=1}^3 \sum_{a=1}^8 
\int \, d^3 r \, \left[\vec r \times \langle p \uparrow \vert 
 \left ( \vec E_i^a (\vec r) \times
\vec B_j^a(\vec r) \right ) \vert p \uparrow \rangle\right]^3\, ,
\label{j2}
\ee
\begin{eqnarray}
\Delta G(\mu^2) & = & \sum_{{{i}},{{j}}=1}^3\sum_{a=1}^8\int{d^3r
\langle p\uparrow|\Bigg\{
\left[\vec{E}_{{i}}^{{a}}({{\vec{r}}})\times
\vec{A}_{{j}}^{{a}}({{\vec{r}}})\right]^3+
\left[\vec{E}_{{i}}^{{a}}({{\vec{r}}})\times
\vec{\nabla}f_{{j}}^{{a}}({{\vec{r}}})\right]^3+} \nonumber \\
&&\mbox{}+\vec{A}_{\perp i}^{{a}}({{\vec{r}}})\cdot
\vec{B}_{\perp j}^{{a}}({{\vec{r}}})
+\vec{\nabla}_\perp f_{{i}}^{{a}}({{\vec{r}}})\cdot
\vec{B}_{\perp j}^{{a}}({{\vec{r}}})
\Bigg\}|p\uparrow\rangle\;,
\label{2bis}
\end{eqnarray}
where $f^a(\vec r) \equiv \int_0^z d\zeta \, \Phi^a (x,y,\zeta)$. 

Upon integration by parts, the second term in eq.~(\ref{2bis}) 
vanishes, whereas the fourth  term equals the first one plus 
a surface integral, 
which also vanishes in 
the model at hand. 

Setting $\vec E_i^a \equiv \frac{\lambda_i^a}{2} \vec E^i$,  
and analogous relations for $\vec A_i^a, \vec B_i^a$, and denoting 
by $\langle \cdot 
\rangle_{\rm c}, \, \langle \cdot \rangle_{\rm osf}$ the color and 
orbital--spin--flavor expectation values, respectively, we get  
\be
J_g(\mu^2) = 
\frac{1}{4} \, 
\sum_{{{i}},{{j}}=1}^3
\sum_{a=1}^{8} 
\langle  \lambda_i^a \lambda_j^a \rangle_{\rm c}
\int d^3 r \, \left[\vec r \times 
\langle p \uparrow \vert \vec E_i(\vec r) \times \vec B_i(\vec r) 
\vert p \uparrow \rangle_{\rm osf}\right]^3\,,
\label{j3}
\ee
\be
\Delta G(\mu^2)  =  
\frac{1}{4} \, 
\sum_{{{i}},{{j}}=1}^3
\sum_{a=1}^{8} 
\langle  \lambda_i^a \lambda_j^a \rangle_{\rm c}
\int d^3r
\langle p\uparrow|\Bigg\{
2\,  \left[\vec{E}_{{i}}({{\vec{r}}})\times
\vec{A}_{{j}}({{\vec{r}}})\right]^3+
\vec{A}_{\perp i}({{\vec{r}}})\cdot
\vec{B}_{\perp j}({{\vec{r}}})
\Bigg\}|p\uparrow\rangle_{\rm osf}\,.
\label{2ter}
\ee

An explicit model calculation of eqs.~(\ref{j3}, \ref{2ter}) 
requires some 
approximation. For instance, one can replace the color fields 
by their expectation values in the nucleon eigenstate. In this case, 
if one includes the self-interaction terms with 
$i=j$, 
$J_g$ and $\Delta G$ turn out to be exactly zero. The reason for this 
is simple. Since the ground state of the nucleon is symmetric 
with respect to the exchange of any pair of quarks, the matrix elements
in the integrals of 
eqs.~(\ref{j3}, \ref{2ter}) do not depend on the particle 
indices $i$ and $j$. Thus 
the integrals 
factorize out 
and $J_g$ and $\Delta G$ vanish exactly because 
$\sum_{i,j} \sum_a \langle \lambda_i^a  
 \lambda_j^a \rangle =0$. In 
\cite{jaffe} the self--interaction terms 
are neglected since they could diverge 
and therefore require to be renormalized. The result
is a negative $\Delta G$, which is a direct consequence of 
$ \sum_a \lambda^a_i \lambda^a_j = -8/3$ for $i \ne j$. 

In our calculation of eqs.~(\ref{j3}, \ref{2ter}) we will include  
also the self--interaction contributions. Let us consider 
for definiteness $\Delta G$ (the calculation 
of $J_g$ proceeds in a similar way).  
We insert in the matrix element of eq.~(\ref{2ter}) a 
complete set of intermediate states $\vert \beta \rangle$, which 
include the orbital excitations of the three--quark system. 
In our non-relativistic calculation only positive energy states
will appear.
The matrix elements
\be
\langle 0|\vec{E}_{{i}}({{\vec{r}}})|{{\beta}}\rangle,\qquad
\langle{{\beta}}|\vec{A}_{{j}}({{\vec{r}}})|0\rangle
\ee
for states
$|\beta\rangle$ with orbital mixed symmetry now carry a 
non trivial dependence on the particle index. 
As a consequence, the integral in eq.~(\ref{2ter}) 
no longer factorizes out
and $\Delta G$ is not forced to vanish, even including the 
self--field terms. The procedure outlined above corresponds
to taking into account
static contributions to the self-spin and is equivalent to
introducing a time-independent Green function for the quarks,
expanded on the modes of the model Hamiltonian.

Eq.~(\ref{2ter}) then reads (we omit the second term 
which turns out to be zero) 
\be
\Delta G(\mu^2)  =  
 \frac 12 \sum_{{{i}},{{j}}=1}^3 
\sum_{a=1}^{8} {{\sum_\beta }}
\left \langle {\lambda}_i^a \cdot
{\lambda}_j^a \right\rangle_{\rm c}
{{\int{d^3r}}}\left[\langle p\uparrow|
\vec{E}_{{i}}({{\vec{r}}})
{{|\beta \rangle}}\times{{\langle \beta|}}
\vec{A}_{{j}}({{\vec{r}}})|p\uparrow\rangle_{\rm osf}\right]^3\,. 
\label{2quater}
\ee
Since the color electric field does not contain any spin--flavor 
operator, the intermediate states $\vert \beta \rangle$ must have 
the same spin--flavor structure as the proton. Moreover,  
the totally symmetric or antisymmetric intermediate states  
give again ($i,j$)--independent 
matrix elements and therefore a vanishing contribution to 
eq.~(\ref{2quater}). 

Before showing our results, we need to discuss the convergence
of the series obtained in the mode expansion. 
The latter is expected to be at worst logarithmically divergent,
as it happens in QED for an isolated electron \cite{jaffe}.
Thus, in principle renormalization is required. On the other hand
our calculation is performed in a non-relativistic effective model,
which is supposed to be valid up to excitation
energies of the order of few
GeV at most. For larger energies relativistic
corrections should be included. The situation here is rather similar
to the one presented in the classical paper of Bethe on the Lamb 
shift \cite{bethe}. In that case a natural cut-off was introduced
and assumed to be equal to the electron mass. Higher energy 
contributions 
were left to be computed using a relativistic approach
and renormalizing. Since the 
divergence was a logarithmical one, the precise value of the cutoff was 
irrelevant and a very good estimate of the energy shift
was obtained. Here we 
adopt the same procedure, taking into account excited states
having energies up to $\sim$ 1 GeV above the ground state. 

We calculated analytically the contributions 
to $\Delta G$ and $J_g$ using the 
IK wavefunctions \cite{IK}. The parameters of the model are fixed 
so as to reproduce the hadronic spectroscopy \cite{IK}. 
The results are shown in Tab.~1, where the contributions 
are labelled 
by the energy quantum number 
$\cal N$ and by the orbital angular momentum $\cal L$ 
of the excited states. They are also classified 
according to the symmetry of the wavefunctions
with respect to the exchange of quark pairs. 
Notice that only the states with mixed symmetry
give a non vanishing contribution, as we anticipated above.

The resulting estimates for $\Delta G$ and $J_g$ are given 
by the sum of the contributions in the last two columns 
of Tab.~1. Only the 
first three energy levels are presented in the 
table. 
For $\Delta G$ we pushed the calculation 
up to the states with 
${\cal N}=3$ and we found 
$0.03 \, \alpha_s$. 
The apparently stronger convergence of the $J_g$ expansion can
perhaps be related to its gauge-invariant expression. 
On the contrary, $\Delta G$ is in general non-local and 
the local expression here adopted is obtained choosing a particular
gauge. It would be interesting to ascertain if, using directly
the non-local expression which introduces a natural cut-off, 
the expansion do converge.

The values of $J_g$ and $\Delta G$ have been expressed 
so far in terms of the 
strong coupling constant $\alpha_s$, which is 
fixed in the IK model 
 by reproducing the $\Delta - N$ mass splitting. 
This gives $\alpha_s=0.9$ \cite{IK}, which  corresponds to
a model scale $\mu_0^2 \simeq 0.25$ GeV$^2$. 
The dynamical quantities 
computed in the model are implicitly 
taken at this scale. 

We can now present our final estimate
\[
J_g (\mu_0^2) \simeq \Delta G(\mu_0^2) \simeq 0.24\,,\;\;\;  
\mu_0^2 = 0.25 \, {\rm GeV}^2.
\]
Thus, at the model scale, 
very little room (if any) is left 
for the orbital angular momentum of gluons: $L_g(\mu_0^2) 
\simeq 0$. 

In order to see what happens at a larger scale, we perform 
a leading--order (LO) QCD evolution. This is of course a dangerous 
thing to do  when the initial scale is 
very low, as in our case.
However, the evolution 
of $J_g$ is known only at LO \cite{hoodbhoy,ji}, 
 and we content ourselves 
with a semiquantitative result. $J_g$ and $\Delta G$
evolve differently \cite{hoodbhoy}. 
At LO the quantity 
$\alpha_s \Delta G$ is constant, whereas $J_g$ increases slowly, 
its asymptotic value being $8 / (16 + 3 N_f) 
= 0.32$. We find 
\be
\Delta G (5 \, {\rm GeV}^2) = 0.70 \,,\;\;\; 
J_g (5 \, {\rm GeV}^2) = 0.31\,.
\nonumber 
\ee
Hence, as $Q^2$ gets larger, the gluon orbital angular 
momentum $L_g$ 
becomes negative and 
increases in absolute value. This scenario is 
similar to the one suggested by the QCD sum rule calculation 
of Ref.~\cite{balitsky}.  Of course, a consistent check of the angular 
momentum sum rule, eq.~(\ref{1}), requires $\Delta \Sigma$ to be 
calculated 
in the presence of 
polarized glue. This calculation is in progress
and will be reported in a future work.

In conclusion, let us summarize our results. 
We showed that the quark model does not 
imply a negative gluon helicity, which would contradict
the indications coming from polarized DIS data. In a
non relativistic model, taking into account self-interaction
contributions, 
we obtained a value of $\Delta G$ 
which is positive and 
close to the range of values prescribed by the 
phenomenological analyses of data. Finally,  
we estimated  $J_g$ and found that it is 
almost equal to $\Delta G$ at the model scale. 

\vspace{0.5cm}

It is a pleasure to thank Z. Berezhiani, L. Caneschi, R. Cenni and 
U. Tambini for many useful discussions.

\pagebreak

\begin{table}
\begin{center}
\begin{tabular}{||c|c|c|c|c||}\hline
& & & &\\
${\cal N}$ & ${\cal L}$ & {\sc Sym} & $\Delta G$ & $J_g$\\ 
& & & &\\
\hline\hline
& & & &\\
0 & 0 & S & 0 & 0 \\ 
& & & &\\ 
\hline
& & & &\\ 
1 & 1 & M & $0.14\;\alpha_s$ & $0.24\;\alpha_s$\\ 
& & & &\\ 
\hline
& & & &\\ 
2 & 0 & S & 0 & 0 \\
2 & 0 & M & $0.08\;\alpha_s$ & $0.02\;\alpha_s$\\ 
& & & &\\ 
2 & 1 & A & 0 & 0 \\ 
& & & &\\ 
2 & 2 & S & 0 & 0 \\
2 & 2 & M & $0.01\;\alpha_s$ & $\sim 0.00\;\alpha_s$\\
& & & &\\ 
\hline
\end{tabular}
\end{center}
\caption{Contributions to $\Delta G$ [eq.~(\protect\ref{2quater})] 
and to $J_g$ from
different intermediate states $|\beta\rangle$. These are classified according 
to their energy and orbital angular momentum numbers, and their symmetry under 
quark pair exchange.}
\end{table}
\end{document}